POLICY FORUM

ARTIFICIAL INTELLIGENCE

# Advancing science- and evidence-based AI policy

Policy must be informed by, but also facilitate the generation of, scientific evidence

Rishi Bommasani[1], Sanjeev Arora[2,3], Jennifer Chayes[4,5,6,7,8] Yejin Choi[1,9], Mariano-Florentino Cuéllar[10], Li Fei-Fei[1,9], Daniel E. Ho[1,9,11,12,13,14], Dan Jurafsky[9,15], Sanmi Koyejo[9], Hima Lakkaraju[16,17], Arvind Narayanan[2,18], Alondra Nelson[19,20], Emma Pierson[6], Joelle Pineau[21], Scott Singer[10,22,23], Gaël Varoquaux[24], Suresh Venkatasubramanian[25,26], Ion Stoica[6], Percy Liang[1,9], Dawn Song[6,27]

Policy-makers around the world are grappling with how to govern increasingly powerful artificial intelligence (AI) technology. Some jurisdictions, like the European Union (EU), have made substantial progress enacting regulations to promote responsible AI. Others, like the administration of US President Donald Trump, have prioritized "enhancing America's dominance in AI." Although these approaches appear to diverge in their fundamental values and objectives, they share a crucial commonality: Effectively steering outcomes for and through AI will require thoughtful, evidence-based policy development (*1*). Though it may seem self-evident that evidence should inform policy, this is far from inevitable in the inherently messy policy process. As a multidisciplinary group of experts on AI policy, we put forward a vision for evidence-based AI policy, aimed at addressing three core questions: (i) How should evidence inform AI policy? (ii) What is the current state of evidence? (iii) How can policy accelerate evidence generation?

AI policy should advance AI innovation by ensuring that its potential benefits are responsibly realized and widely shared. To achieve this, AI policy-making should place a premium on evidence: Scientific understanding and systematic analysis should inform policy, and policy should accelerate evidence generation. But policy outcomes reflect institutional constraints, political dynamics, electoral pressures, stakeholder interests, media environment, economic considerations, cultural contexts, and leadership perspectives. Adding to this complexity is the reality that the broad reach of AI may mean that evidence and policy are misaligned: Although some evidence and policy squarely address AI, much more partially intersects with AI. Well-designed policy should integrate evidence that reflects scientific understanding rather than hype (*2*). An increasing number of efforts address this problem by often either (i) contributing research into the risks of AI and their effective mitigation or (ii) advocating for policy to address these risks. This paper tackles the hard problem of how to optimize the relationship between evidence and policy (*3*) to address the opportunities and challenges of increasingly powerful AI.

## HOW SHOULD EVIDENCE INFORM POLICY?

Developing evidence-based policy is critical for AI, especially considering its role within the extant policy ecosystem. For example, in the US, the Trump administration should leverage the Foundations for Evidence-Based Policymaking Act (Evidence Act) that the president signed in 2019 during his first term. The bipartisan Evidence Act requires federal agencies to provide evidence-building plans to justify funding, share nonsensitive data to enable research, and designate a chief evaluation officer to oversee agency-wide evaluation of program effectiveness.

Defining what counts as (credible) evidence is the first hurdle for applying evidence-based policy to an AI context—a task made more critical given that norms for evidence vary across policy domains. In health policy, evidence generally refers to randomized control trial results or observational data. In economic policy, evidence tends to be more expansive, encompassing theoretical approaches (e.g., macroeconomic forecasts) alongside data-driven approaches (*4*).

To advance positive public outcomes, evidence-based policy requires careful execution that properly aligns incentives. Historically, evidence-based policy has at times been co-opted to justify inaction (*3*) or promote negative societal outcomes (*5*). The tobacco industry relied on inconclusive studies to play up uncertainty, inhibiting policy to address documented tobacco-related harms (*6*). Fossil fuel companies misled the public about climate change despite company-internal reports that anticipated severe harms (*7*). To avoid repeating these failures, evidence-based AI policy would benefit from evidence that is not only credible but also actionable. A focus on marginal risk (*8*), meaning the additional risks posed by AI compared to existing technologies like internet search engines, will help identify new risks and how to appropriately intervene to address them.

## WHAT IS THE STATE OF THE EVIDENCE?

Although many mechanisms contribute to the evidence base on AI capabilities, risks, and impacts, processes for evidence certification (i.e., determining if evidence is credible) and evidence synthesis (i.e., reviewing multiple, possibly conflicting, pieces of evidence) are nascent. In mature policy domains, evidence certification can involve specific systems [e.g., the Grading of Recommendations Assessment, Development, and Evaluation (GRADE) system in health policy] or available proxies (e.g., scientific peer review).

The International Scientific Report on the Safety of Advanced AI (*9*), led by Turing Award winner Yoshua Bengio, is a recent initiative that is well-positioned to certify and synthesize evidence for global AI governance. Inspired by the Intergovernmental Panel on Climate Change (IPCC), the report is authored by 96 AI experts, including an international Expert Advisory Panel nominated by 30 countries, the Organisation for Economic Co-operation and Development (OECD), the EU, and the United Nations (UN). To maintain independence, the experts retain full discretion over the content and are not affiliated with industry. On certification, the report does not rely on peer review exclusively, instead relying on the judgment of the authors to identify high-quality sources based on the indicators of (i) originality, (ii) impact, and (iii) transparency on methods, prior work, limitations, and opposing views.

The International Scientific Report defines three general risk categories: malicious use risks (e.g., cloned voices used in financial scams and biological attacks), risks from malfunctions (e.g., reliability issues where models may generate false content and bias against certain groups), and systemic risks (e.g., labor market disruption and copyright infringement). Given this taxonomy of risks, the International Scientific Report indicates that several risks cause harm today, specifically scams, nonconsensual intimate imagery, child sexual abuse material, reliability issues, bias, and privacy violations. Other risks have partial (though possibly increasing) evidence, specifically large-scale labor market impacts, AI-enabled hacking or biological attacks, and loss of control. As a result, experts arrive at very different predictions for this second category of risks. Expert disagreement on this category of risks can serve as a prompt to policy-makers to explore how they can accelerate evidence generation.

## HOW CAN POLICY ACCELERATE EVIDENCE GENERATION?

As the International Scientific Report notes, policy-makers often make critical AI policy decisions with limited scientific evidence, posing an "evidence dilemma": Acting preemptively might lead to ineffective or unnecessary measures, whereas waiting for stronger evidence could leave society vulnerable. In response, policy-makers should not remain idle: Policy can actively accelerate the generation of evidence that can best inform future policy decisions. We propose specific mechanisms that policy-makers should pursue to grow the evidence base and serve as the foundation of evidence-based AI policy.

### Incentivize pre-release evaluation

Prior to releasing a model, model developers and external parties can evaluate the model to better understand how its release will affect society. Model developers should proactively measure risks prior to deployment; such evaluations can clarify the extent to which models pose marginal risks. Currently, many leading developers have signed onto the Frontier AI Safety commitments, which include a commitment to pre-release evaluation. Many of these developers (e.g., Anthropic, Google, Meta, OpenAI) do report evaluation results for some risks like bioweapons. These evaluations help clarify that, for example, recent model releases from Anthropic and OpenAI have increased their self-assessed level of risk in the domain. However, recent media reporting suggests that the quality of these evaluations may be degrading [e.g., the time afforded to internal evaluation given intense pressure to rapidly release models (10)].

We recommend that policy-makers incentivize the evaluation of models prior to release. To supplement their internal testing, developers should work with trusted external entities. Following the original initiative from the UK AI Security Institute in 2023, several countries are consolidating state-backed institutes to evaluate advanced AI. Notably, the US still backs science- and evidence-based inquiry at the Center for AI Standards and Innovation (CAISI). For external pre-release evaluation to be most valuable, the resulting evidence should be correctly interpreted: Model developers and external testers should publicly clarify the conditions involved in testing (e.g., if evaluators are paid by developers or are unable to disclose unfavorable results; the time and level of access afforded to evaluators; or how the developers used the evaluation results to inform release decisions).

### Increase information sharing

Pre-release capability and safety evaluations provide insight into model capabilities and risks but are alone inadequate for assessing the societal impact of AI. More generally, key information about AI and its societal impact is siloed within AI companies. For example, the 2024 Foundation Model Transparency Index scores 14 major AI companies (e.g., Anthropic, Google, Microsoft, Meta, OpenAI) for their transparency on a range of issues; most companies score poorly for publicly sharing information on how they mitigate risk. In light of these evidence gaps, policy-makers in multiple jurisdictions have enacted transparency requirements. Most policies, such as the EU AI Act, focus on sharing safety-related information specifically with governments. Although this approach is valuable and there are many valid reasons to restrict information sharing to trusted actors, public transparency is essential for true accountability. In practice, it is often citizens, journalists, civil society organizations, and academics who are at the forefront of identifying sociotechnical harm (11).

We recommend that policy-makers require major AI companies to disclose more information about their safety practices to governments and, especially, to the public. First, transparency obligations should reflect informational needs: Disclosing safety frameworks clarifies the steps that developers take internally to mitigate risk, and disclosing evaluation results clarifies the current level of measured risk. Second, transparency obligations should reflect who will best use information: Given the broad network of entities responsible for advancing accountability (11), policies should prioritize information sharing with the public, and not just governments, in many cases. Finally, transparency obligations should be proportionate to not impose undue burdens on developers: Criteria used to differentiate obligations should have specified processes for how they will update and avoid absolute reliance on fraught proxies such as training-time compute or monetary costs (12).

### Monitor postdeployment impacts

Once AI systems are deployed, especially at scale, they affect society in a variety of ways. Yet surprisingly little is known about these impacts: The 2024 Foundation Model Transparency Index highlights that, of all issues, major AI companies are the least transparent about their models' postdeployment adoption. Yet, the patterns of adoption clarify how general-purpose technologies like today's AI models shape society in specific ways. Recently, Anthropic has built an Economic Index, which reports statistics about how Anthropic's models are used based on clustering user queries to make progress on understanding effects of AI on the labor market and broader economy. Beyond the efforts of individual companies, adverse event–reporting databases, such as those proposed by the US National AI Advisory Committee, are critical to grow the collective evidence base by documenting concrete instances of harm in practice. Although initial attempts like the AI Incidents Monitor from the OECD provide coverage of adverse events, precise standards for (i) which entities are responsible for reporting, (ii) what constitutes an adverse event, and (iii) which parties need to be informed of an adverse event do not yet exist.

We recommend that policy-makers increase postdeployment monitoring of AI harms. Related domains, like cybersecurity, can guide how to design and implement postdeploying harm monitoring for AI. For example, the US Cybersecurity and Infrastructure Security Agency administers an incident-reporting database under the Cyber Incident Reporting for Critical Infrastructure Act of 2022. Initiatives like this successfully address challenges that arise in AI such as how to coordinate disclosure of a vulnerability to many affected model developers and system providers, especially given that issues with AI models and systems may generalize across different models (13).

### Protect third-party research

Important information about AI is often siloed within companies that develop and deploy the technology, but relevant expertise is more widely distributed. Third-party research by independent parties is an indispensable form of evidence, given the independence from commercial incentives and, thereby, the greater trust it may confer. To conduct research on released AI models and deployed AI systems, third-party researchers require access to these technologies. Although most major foundation models are available to researchers in some form, usually through an API (e.g., OpenAI's GPT-4o) or through their weights (e.g., Meta's Llama 3.3), current practices that govern access inhibit third-party research. The terms of service for many leading developers include clauses that, potentially inadvertently, suppress third-party research. For example, to rigorously measure misuse risk, researchers often jailbreak a model to circumvent the model's safeguard to refuse malicious requests, which violates standard terms of service, in turn risking platform bans or legal action.

[1]Stanford Institute for Human-Centered Artificial Intelligence, Stanford University, Stanford, CA, USA. [2]Department of Computer Science, Princeton University, Princeton, NJ, USA. [3]Princeton Language and Intelligence, Princeton University, Princeton, NJ, USA. [4]College of Computing, Data Science, and Society, University of California, Berkeley, Berkeley, CA, USA. [5]School of Information, University of California, Berkeley, Berkeley, CA, USA. [6]Department of Electrical Engineering and Computer Sciences, University of California, Berkeley, Berkeley, CA, USA. [7]Department of Mathematics, University of California, Berkeley, Berkeley, CA, USA. [8]Department of Statistics, University of California, Berkeley, Berkeley, CA, USA. [9]Department of Computer Science, Stanford University, Stanford, CA, USA. [10]Carnegie Endowment for International Peace, Washington, DC, USA. [11]RegLab, Stanford University, Stanford, CA, USA. [12]Stanford Law School, Stanford University, Stanford, CA, USA. [13]Department of Political Science, Stanford University, Stanford, CA, USA. [14]Stanford Institute for Economic Policy Research, Stanford University, Stanford, CA, USA. [15]Department of Linguistics, Stanford University, Stanford, CA, USA. [16]Harvard Business School, Harvard University, Boston, MA, USA. [17]Department of Computer Science, Harvard University, Boston, MA, USA. [18]Center for Information Technology Policy, Princeton University, Princeton, NJ, USA. [19]School of Social Science, Institute for Advanced Study, Princeton, NJ, USA. [20]Science, Technology, and Social Values Lab, Institute for Advanced Study, Princeton, NJ, USA. [21]School of Computer Science, McGill University, Montréal, QC, Canada. [22]Oxford China Policy Lab, Oxford, UK. [23]Oxford Martin AI Governance Initiative, University of Oxford, Oxford, UK. [24]Inria Saclay – Île-de-France, Palaiseau, France. [25]Department of Computer Science, Brown University, Providence, RI, USA. [26]Data Science Institute, Brown University, Providence, RI, USA. [27]Berkeley Center for Responsible Decentralized Intelligence, Berkeley, CA, USA. Email: nlprishi@stanford.edu

We recommend that policy-makers create shields to protect good-faith third-party AI research. In May 2024, 350 leading AI researchers and advocates signed an open letter that advocated for a safe harbor to protect such research (14). The proposed safe harbor draws inspiration from cybersecurity, where similar safe harbor provisions exist at the federal level in the United States (15). In general, if researchers follow established rules of conduct and responsibly disclose issues with AI technologies to advance the public interest, then indemnification from legal liability would create the appropriate incentives to grow the evidence base.

### Prioritize well-evidenced interventions

Successfully mitigating the risks of AI requires sociotechnical approaches that recognize the role of humans, organizations, and technology in a defense-in-depth approach. Defense-in-depth means layering technical system-level interventions with a broader suite of societal interventions elsewhere in the supply chain (e.g., in a bioweapons context, one might focus not only on an AI system used to design a weapon but also on vendors of biological materials needed to synthesize the weapon). Critically, for different risks, the evidence regarding potential interventions at specific points in the supply chain should inform where policy targets interventions.

We recommend that policy-makers strengthen societal defenses, especially given clear evidence of unmitigated risk even absent AI capabilities. For many threat vectors surrounding malicious use, AI capabilities are exploited as an intermediary step in a more complex process (e.g., synthesizing disinformation that is then disseminated through social media networks, obtaining information that is then used to build bioweapons). To address potential marginal risk from AI for these threat vectors, downstream interventions may be effective while also reducing preexisting non-AI risk for the same threat vectors. For example, prevalent cybersecurity practices are insufficient according to experts: Hardening software systems would not only address these long-standing issues but also better prepare society for AI-related cyberattacks.

### Bridge fragmented subcommunities

Currently, the AI community is fractured, with many divergent views on how to approach risk and policy. At present, the degree of consensus or the lack thereof remains unclear for many critical questions (e.g., what forms of evidence are credible for supporting the claim that frontier models pose marginal risk in enabling bioweapons development?). Forging consensus will be difficult given the striking divides in the AI community on core issues (e.g., the rate of technological progress). With this in mind, the types of deliberative processes that facilitate consensus formation among experts in other more mature policy domains, to our knowledge, have not been systematically explored in the context of AI.

We recommend that policy-makers catalyze the formation of scientific consensus. Scientific consensus, including on areas of uncertainty or immaturity, is a powerful primitive for better AI policy. Existing efforts such as the previously mentioned International Scientific Report, UN High-level Advisory Body on Artificial Intelligence (HLAB-AI), and global network of AI Safety Institutes are key initial steps to build international consensus. Close policy-maker–scientist partnerships can accelerate this process, and other industries provide useful historical precedent. For example, HLAB-AI recommends forming an International Scientific Panel on AI, which could prepare consensus reports akin to the efforts of the IPCC. Efforts to form scientific consensus should incorporate core principles identified in other domains, namely that these processes be seen as legitimate within the community of experts (i.e., involve those with demonstrable expertise and be broadly inclusive) and credible to the external ecosystem (i.e., involve experts who maintain independence from undue coercion).

### 21ST-CENTURY GRAND CHALLENGE

To advance evidence-based AI policy, the evidence-generating mechanisms must be complemented by efforts to strengthen the underlying scientific inquiry into AI risks and how to successfully identify and mitigate them. For example, alongside improving our understanding of risks due to current technology, we should simultaneously invest in long-term research to develop inherently safer AI. More generally, while we focus on policies that can grow the evidence base, future work should focus on the relationship between evidence and action. If-then protocols that map evidence to policy responses are adopted in many mature policy domains (e.g., fiscal policy, climate policy, disaster policy). This approach can allow parties that disagree about when specific evidence will materialize to, instead, productively collaborate on the response.

The effective governance of AI is a grand challenge for the 21st century. Evidence from currently siloed disciplines is foundational to effective governance, and accelerating evidence building is foundational to policy that evolves in concert with technology. However, we recognize that coalescing around an evidence-based approach is only the first step in reconciling many core tensions. Extensive debate is both healthy and necessary for democratically legitimate policy-making; such debate should be grounded in available evidence. ■

**ACKNOWLEDGMENTS**

We thank A. Madry, H. Toner, K. Klyman, M. Brundage, P. Henderson, S. Kapoor, and Y. Bengio for helpful discussion. S.A. is employed part-time at Meta; Y.C. is a consultant to NVIDIA and previously served as a consultant to Ai2; M.-F.C. is an advisory board member of the Stanford Institute for Human Centered Artificial Intelligence, Visiting Scholar at Stanford Law School, and a board member of Inflection AI; L.F.-F. is at World Labs LLC; S.K. is a cofounder of Virtue AI; A.Ne. is on the board of directors of Mozilla, the Advisory Council of the Institute for Ethics in AI, University of Oxford, and the Advisory Board of Data and Society; S.V. is on the boards of Data and Society, and the Partnership on AI; I.S. is the executive chairman at Databricks and Anyscale; P.L. is at Together AI, Simile AI, Arell AI, and Virtue AI and is an adviser to Snowcrash and Humanitas; and D.S. is founder of Oasis Labs and is at Virtue AI. This effort represents scholarly work from each contributor in their personal capacity. Each author is identified by their primary academic institutional affiliation, though several authors hold additional affiliations. In all cases, this piece solely reflects the views of the authors and does not represent the positions of their affiliated organizations.

10.1126/science.adu8449